\documentclass[11pt, oneside]{article}   	
\usepackage{geometry}                		
\usepackage{graphicx}				
\usepackage{amssymb}
\usepackage{url}
\usepackage{color}
\usepackage{natbib}
\usepackage{setspace}
\usepackage{lineno}

\begin{document}
\noindent
\textbf{A Composite Genome Approach to Identify Phylogenetically Informative Data from Next-Generation Sequencing}

Rachel S. Schwartz,$^{\ast,1}$ Kelly Harkins,$^{2}$ Anne C. Stone,$^{2}$ and Reed A. Cartwright,$^{1,3}$

$^{1}$The Biodesign Institute

$^{2}$School of Human Evolution and Social Change

$^{3}$School of Life Sciences

Arizona State University, Tempe, AZ, USA

E-mail: Rachel.Schwartz@asu.edu

\abstract{
We have developed a novel method to rapidly obtain homologous genomic data for phylogenetics directly from next-generation sequencing reads without the use of a reference genome.
This software, called SISRS, avoids the time consuming steps of \textit{de novo} whole genome assembly, genome-genome alignment, and annotation.
For simulations SISRS is able to identify large numbers of loci containing variable sites with phylogenetic signal.
For genomic data from apes, SISRS identified thousands of variable sites, from which we produced an accurate phylogeny. 
Finally, we used SISRS to identify phylogenetic markers that we used to estimate the phylogeny of placental mammals. 
We recovered phylogenies from multiple datasets that were consistent with previous conflicting estimates of the relationships among mammals.
SISRS is open source and freely available at https://github.com/rachelss/SISRS.
}


\section*{{Introduction}\label{sec:Intro}}

Until recently, phylogenetic studies relied on tens of loci (at most) from the genome to determine evolutionary relationships \citep{Giribet2001,Harpke2013}. 
However, these datasets often had insufficient information to provide strong support for all the relationships of interest \citep{Stanley2011}. 
Recent improvements in sequencing technology have enabled phylogenetic studies to use larger datasets in an attempt to resolve previously undetermined or controversial evolutionary relationships, but the area is still in its infancy \citep{Burleigh2011,Cohen2012,Crawford2012,Delsuc2005,Faircloth2012,McCormack2012,Mccormack2013b,Yoder2013}. 

There are currently several approaches to producing large datasets for phylogenetics. 
First, whole genomes are sequenced and assembled; genomes are then compared to identify homologous regions for phylogenetics \citep{Yoder2013}.
The drawback of this approach is the time required to construct quality assemblies and identify homologous regions, either by annotating the genome or using genome comparison tools.
Furthermore, because distantly related taxa may not be easily comparable, phylogenetic analyses using whole-genome comparisons have focused on closely related species for which alignments are possible \citep{Yoder2013,Fan2013}.

In a second approach, shotgun sequence data are aligned to a reference genome.
This method assumes a reference genome, which is not always available.
As with whole genome comparisons, the \textit{de novo} assembly of a high-quality genome to be used as a reference requires high-coverage data and significant time.
However, even given a reference genome, homologous loci may not be recoverable for species distantly related to the reference \citep{Bertels2014}.

Other approaches involve sequencing a subset of the genome.
One such approach screens existing datasets for variation in the taxa of interest \citep{ONeill2013, Senn2013, Steele2008}.
In another, regions that are conserved across taxa are identified from whole-genome alignments; both the conserved elements and regions adjacent to them may contain phylogenetic information \cite {Crawford2012,Faircloth2012,McCormack2012,Lemmon2012}. 
In a third, a consistent subsample of the genome may be sequenced \cite[e.g.\ using RADseq;][]{Eaton2013}.
However, the drawback of these approaches is that new phylogenetic markers must be developed for each research study; significant time is often required for marker development and these data have limited potential for reuse. 
Additionally, a consistent, phylogenetically informative subsample of the genome may be difficult to obtain at deep taxonomic levels.

Here we describe a novel computational tool, SISRS (pronounced ``scissors''), to identify informative data for phylogenetic studies directly from shotgun sequencing of whole genomes. 
SISRS, which stands for Site Identification from Short Read Sequences requires neither a reference genome nor \textit{a priori} knowledge of potentially informative loci. 
Our software circumvents the difficulties in identifying homologous loci from whole-genome alignments when rearrangements have occurred because the conserved regions are not required to share identifiable synteny across taxa. 
SISRS also takes advantage of the raw data to avoid erroneous genotype calling due to sequencing error and copy number variable regions (CNVs). 

\begin{figure}[!ht]
\includegraphics[trim=0cm 10cm 0cm 0cm, clip=true, width=\textwidth]{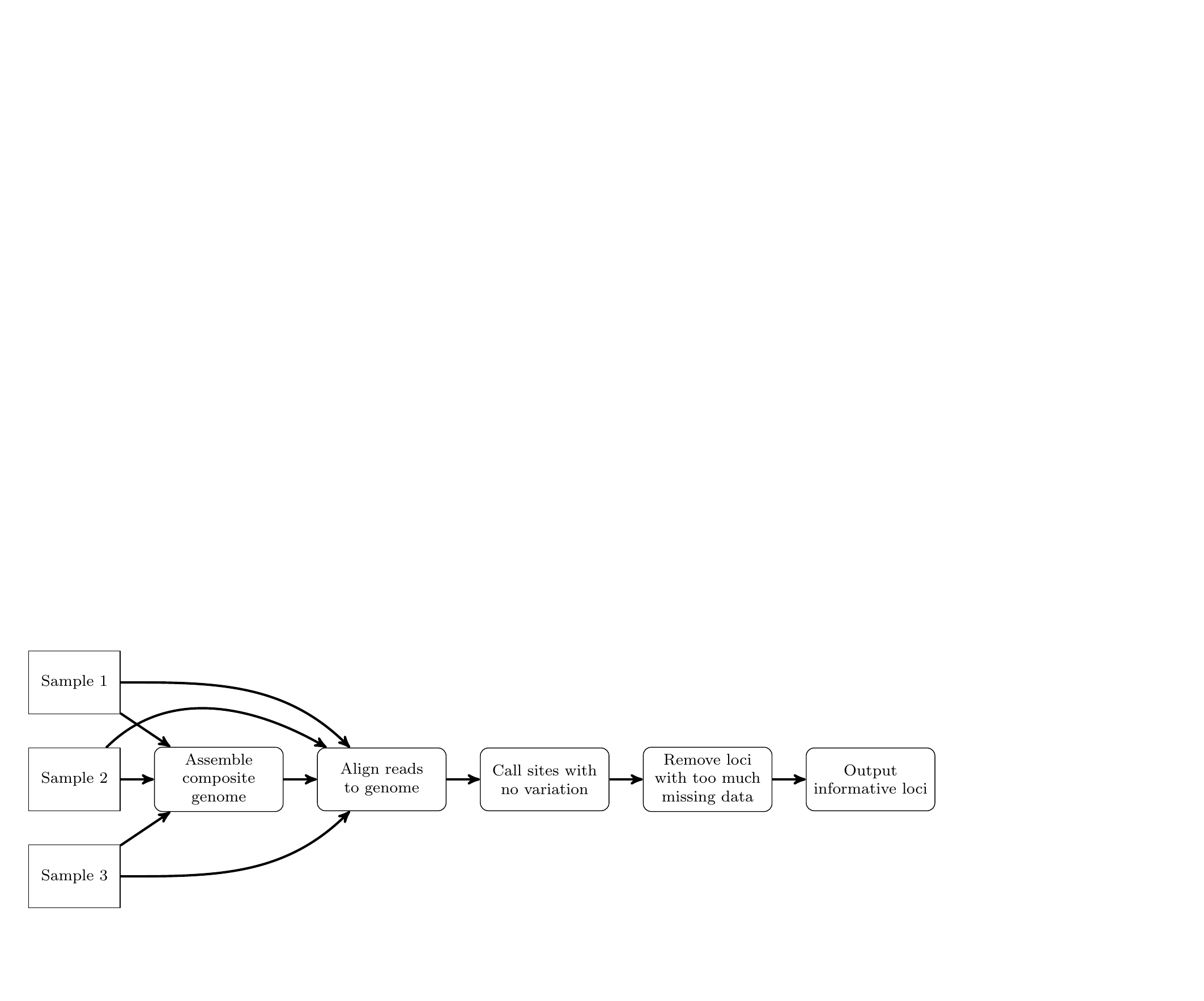}
\caption{SISRS framework to identify phylogenetically informative sites.}
\label{fig:pipeline}
\end{figure}

SISRS identifies phylogenetically informative regions via a novel protocol. 
(1) SISRS assembles a ``composite genome'' from shotgun sequencing reads for all taxa.
(2) The composite genome is used as a reference to align the sequencing data for each sample. 
(3) The sequence for each sample is identified via a strict consensus (i.e.\ sites that are variable are called as unknown). 
(4) SISRS removes loci that have too much missing information (as specified by the user).
In this way, SISRS identifies sites across entire genomes that are phylogenetically informative and reduces errors due to biological and experimental error.
This protocol is outlined succinctly in Figure \ref{fig:pipeline}. 

We demonstrate that SISRS provides high quality phylogenetic datasets across a range of simulated and empirical data. 
First, the data output by SISRS for simulated shotgun reads was congruent with the starting phylogeny at all depths in the tree. 
Second, using previously sequenced shotgun data for seven primate taxa, we were able to rapidly identify homologous data using SISRS and estimate the known phylogeny accurately. 
Third, we used available data to estimate the phylogeny of mammals, the root of which has remained controversial.
Using SISRS, phylogenies can be produced from next-generation sequencing reads in a matter of days.
For example, identification of phylogenetic markers from raw mammalian transcriptome reads took just 15 hours.

\section*{Methods\label{sec:Methods}}
\subsection*{Composite reference genome}

SISRS currently uses a de Bruijn-graph de novo assembler, Velvet \citep{Zerbino2008}, to construct a composite reference genome.
Analysis of alternative algorithms is ongoing.
The composite genome is assembled from a subset of raw NGS reads (i.e.\ FASTQ files) from multiple taxa.
When using a subset of the data, conserved regions are likely to be assembled.
Regions of the genome that are either unique to a single taxon or highly variable among taxa are less likely to be included in the composite genome. 
To construct the subset, SISRS uses reservoir sampling \citep{Vitter1985} from the data for each species.
The size of the subsample is determined based on a user-specified genome size such that the subsample includes approximately 10x coverage of conserved regions across species.
After assembling these diverse data, the resulting contigs contain composite sequences of loci that are conserved across some or all taxa

\subsection*{Site calling for each position in the reference}
To determine conserved regions and variation among species within these regions, the full sequencing data is mapped back to the composite reference genome using Bowtie 2 \citep{Langmead2012}.
SISRS uses a strict consensus to call the genotype of each species for each site in the composite reference.
This is a conservative approach to eliminate any sites that contain paralogous data (i.e.\ false positive variable sites), thus significantly reducing non-phylogenetic signal \citep{Philippe2011}. 
After genotype calling, SISRS produces a dataset containing sites that have information for most or all taxa (as specified by the user).
To reduce the final dataset, SISRS can produce a final alignment containing only sites that are variable among taxa; this type of data is used in all analyses described below.


\subsection*{Simulations to test methodology}

To determine how well our approach identified phylogenetically informative sites, we simulated 168 datasets of next-generation sequencing reads with different levels of sequencing coverage on multiple phylogenies.
We used four phylogenies to simulate genomes: one two-taxon tree and three eight-taxon trees.
The eight-taxon trees were pectinate with (1) equal internode branch lengths, (2) increasing internode branch lengths from root to tip, and (3) decreasing internode branch lengths from root to tip (Suppl.\ Figure \ref{fig:trees}).
We produced a total of 24 trees with increasing levels of divergence among species by multiplying each branch length by values from 0.01 to 0.06.
We simulated genomes of one million nucleotides on each of these trees using the Jukes-Cantor model with Dawg 2.0 \citep{Cartwright2005,Jukes1969}. 
Illumina-like NGS data were simulated using the software ART (version BananaPancakes-04-02-2013) with an error model that was derived from an empirical dataset produced using an Illumina HiSeq 2000 \citep{Huang2012}.  
Each simulation had either 1, 2, 4, 8, 10, 20, and 50x coverage; reads were 99 bp and paired-end (the maximum allowed by the software given empirically derived parameters).
For each data set we recorded (1) the total number of variable sites simulated, (2) the total number of variable sites output by SISRS, (3) the number of these sites that could be mapped, and (4) the number of mapped variable sites that were true variants.

We expected some variable sites not to be recovered by SISRS and
counted the number of these false negatives due to (1) insufficient data in the simulations (defined as fewer than three reads for that site), (2) lack of coverage by the composite genome (for sites with sufficient data), (3) additional insufficient data due to lack of mapped reads, and (4) additional sites that were filtered out due to variation.
Additionally we counted the number of potential and actual false positive variable sites.
Potential false positives were due to sites that would be called as variable (due to error) based on the simulated reads.
Actual false positives were (1) the subset of these sites that were identified as variable by SISRS, or (2) additional sites due to erroneous read mapping.
To determine the value of the data output by SISRS we counted the number (and percent) of sites that were concordant with the true tree as a function of branch scale, coverage, depth, and tree topology.

\subsection*{Empirical Data Test}

We further tested our approach using data from apes, including human (\textit{Homo sapiens}), chimpanzee (\textit{Pan troglodytes}), bonobo (\textit{P. paniscus}), gorilla (\textit{Gorilla gorilla} and \textit{G. beringei}), and orangutan (\textit{Pongo pygmaeus} and \textit{P. abelii}). 
The crab-eating macaque (\textit{Macaca fascicularis}) and rhesus macaque (\textit{M. mulatta}) were used to root the tree. 
These primates were chosen to test the efficiency and effectiveness of this method on empirical data due to their well-established phylogeny \citep{Perelman2011}. 

Raw Illumina paired-end sequence data was obtained from the European Nucleotide Archive (Suppl.\ Table \ref{table:ENAdata}). 
To reduce the size of the dataset being analyzed, we aligned the data to the human genome (build 37) using Bowtie 2 as in SISRS. 
We extracted only the data that aligned to human chromosome 21. 
These reads were then placed in FASTQ files as new paired-end datasets, as would be generated directly from a sequencing run. 
Potentially informative sites were obtained using SISRS. 
The genome size specified for the composite genome subsampling procedure was 48 million, approximately the size of human chromosome 21.

Due to the size of the dataset, we treated the data output by SISRS as a single concatenated locus \citep{Yoder2013} and 
analyzed the data in a maximum likelihood (ML) framework with 1000 bootstraps implemented in RAxML-HPC2 8.0.3. \citep{Stamatakis2006}. 
The GTRGAMMA model was used; the omission of invariable sites was accommodated with the ASC parameter as recommended in the manual \citep{Lewis2001}.

\subsection*{Estimating the Mammal Phylogeny}

We demonstrate the value of the SISRS approach using 30 placental mammal taxa.
Transcriptome and genome data were obtained as above (Suppl.\ Table \ref{table:ENAdata}).
For most taxa we combined data from more than one individual to increase genome coverage.
Potentially informative sites were obtained using SISRS; however, the composite genome was derived exclusively from transcriptome data to reduce assembly time and memory requirements.
The genome size specified for the composite genome subsampling procedure was 100 million.
The dataset output by SISRS for mammals was analyzed using the same method as the ape dataset.

\section*{Results\label{sec:Results}}

\subsection*{Recovery of phylogenetically informative sites}
\begin{figure}[!ht]
\includegraphics[width=\textwidth]{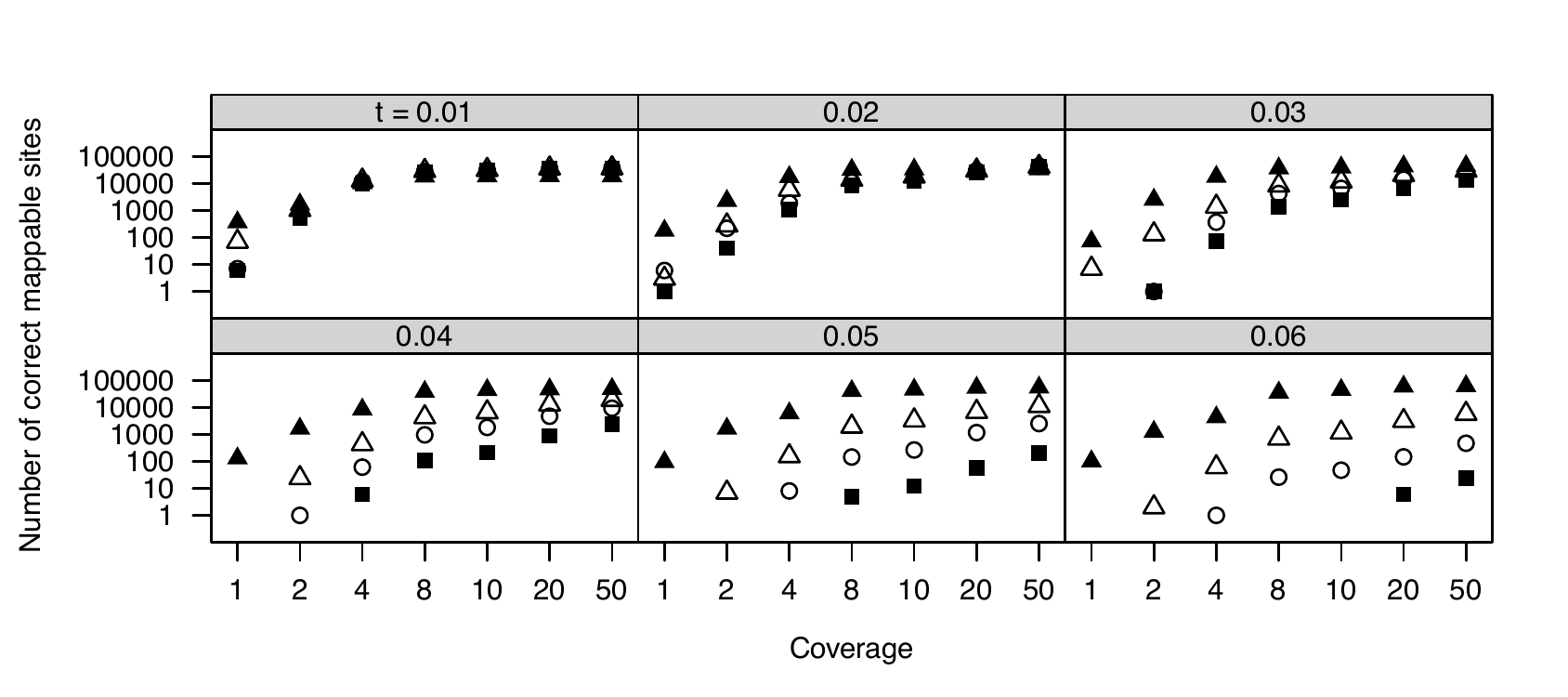}
\caption{SISRS produced substantial amounts of informative data for phylogenies of different shapes and evolutionary distances. 
The number of true variable sites identified from simulated data is shown for each of four trees (symbols) (Suppl.\ Fig. \ref{fig:trees}) for increasing numbers of substitutions between species (panels), and increasing levels of coverage (x axis). Symbols are 
$\blacktriangle$ for two taxon trees; $\circ$ for equal branch length trees; $\blacksquare$ for trees with short deep branches; $\triangle$ for trees with long deep branches.}
These sites were identified from 1 million base pair genomes; thus, larger genomes are expected to produce more sites, particularly as long as a fraction of those genomes are reasonably conserved.
\label{fig:simresults}
\end{figure}

\begin{figure}[!ht]
\includegraphics[width=\textwidth]{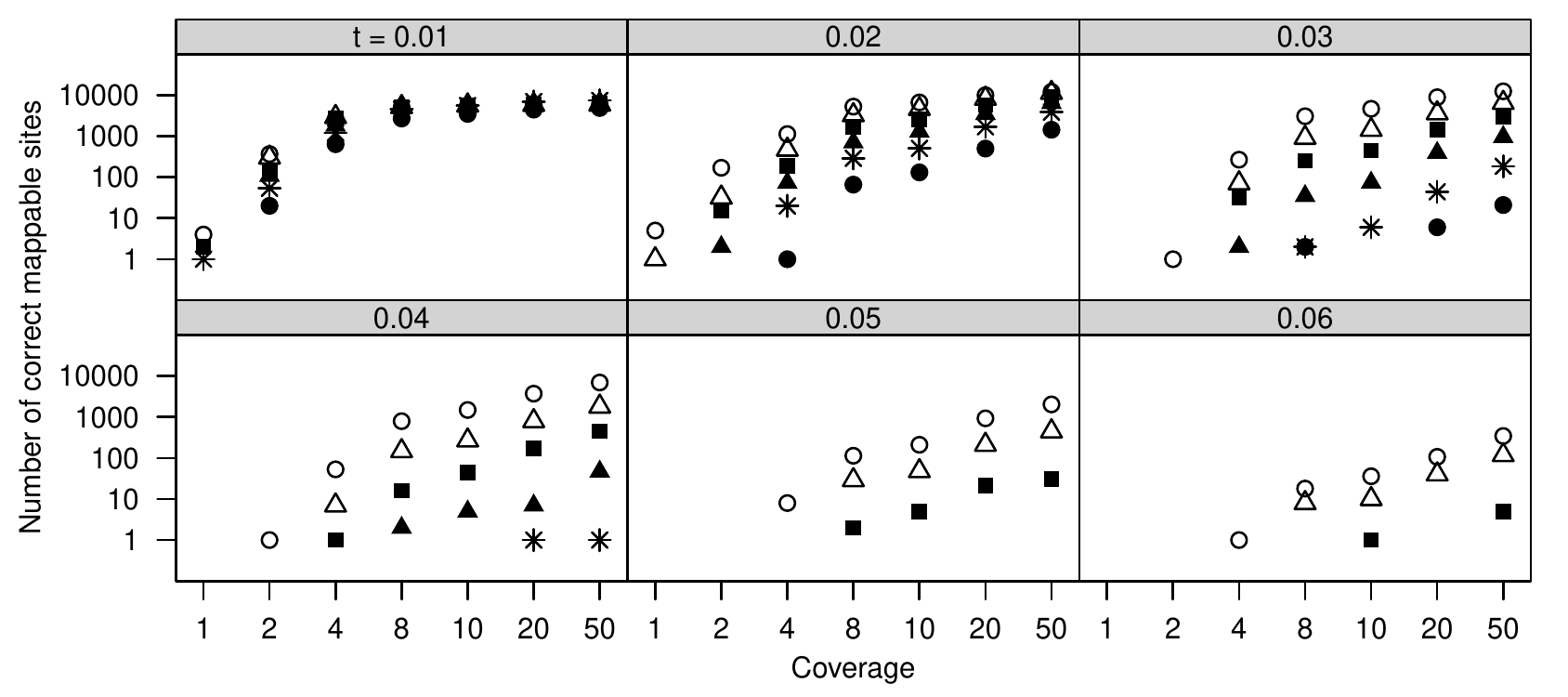}
\caption{SISRS identified fewer informative sites for deeper nodes in the tree; however, some sites identified each split in all cases except for extremely diverged species with low-coverage data). 
Results are separated by branch length (panels), coverage (x axis), and tree depth (symbols). 
The number of sites supporting the node A+B are denoted as $\circ$; $\triangle$ denotes sites supporting A+B+C;$\blacksquare$ denotes A+B+C+D; $\blacktriangle$ denotes A+B+C+D+E; $\ast$ denotes A+B+C+D+E+F; $\bullet$ denotes A+B+C+D+E+F+G.
Only results for the equal-branch-length tree are shown; results for the three eight-taxon trees were similar (Suppl.\ Fig. \ref{fig:sim_results_all_depths}).
Fewer sites were recovered for the tree with short deep branches, compared to the equal-branch-length tree, while more sites were recovered for the tree with long branches.}
\label{fig:simresults_depths}
\end{figure}

\begin{figure}[!ht]
\includegraphics[width=1\textwidth]{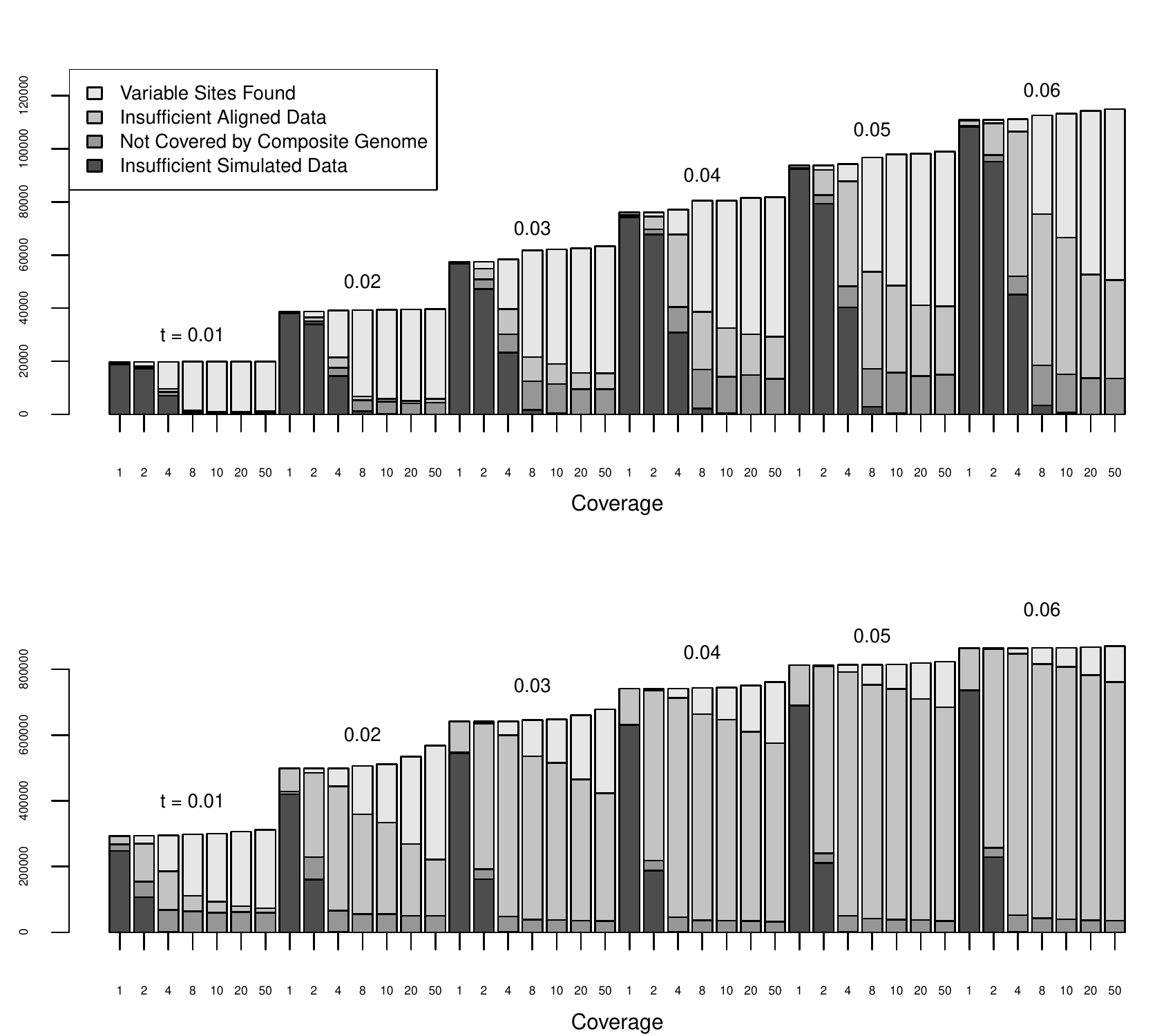}
\caption{SISRS failed to identify some variable sites in simulated data due to low coverage at those sites or lack of coverage by the composite genome.
Top: Results for the two-taxon tree.
Bottom: Results for the eight-taxon tree with equal-length branches.
The results for the other eight-taxon trees are nearly identical.}
\label{fig:sims_falsenegs}
\end{figure}

To determine how well our approach identified phylogenetically informative data, we
simulated genomes on four phylogenies, simulated NGS data on these genomes, and examined how well we were able to recover
variable sites. 
For all simulation trees, the number of potentially informative sites identified using SISRS increased with increased coverage (Figure \ref{fig:simresults}). 
As the distance between taxa (i.e.\ branch length) increased for a given tree, the number of output sites decreased (Figure \ref{fig:simresults}). 
Of these sites, the greatest number allowed the accurate identification of the shallowest nodes within a tree, with a decreasing number allowing the identification of deeper nodes (Figure \ref{fig:simresults_depths}). 
Insufficient coverage in the simulations, lack of coverage by the composite genome, and insufficient coverage following mapping of the reads to the composite genome reference contributed to a failure to recover all variable sites (Figure \ref{fig:sims_falsenegs}). 
Potential false positives due to read simulation error were removed by SISRS as part of calling by strict consensus; however, some new false positives were introduced, likely due to erroneous mapping of reads to the composite genome.
These false positives represented less that 2.5\% of identified variable sites for the two-taxon trees, and 0.2\% for the eight-taxon trees.

\subsection*{Ape tree is recovered}

We also tested the utility of SISRS using NGS data from apes, for which the phylogeny is well established.
This analysis was conducted using 14 cores on a FreeBSD 10.0 server; the total time to produce an alignment from raw reads was 36 hours.
The maximum amount of memory required during the composite genome assembly was 1.5Gb.
We identified 148,639 variable sites that contained observations at least five samples. 
The ML estimate of the phylogeny with 1000 bootstraps was fully concordant with the known phylogeny of apes with all nodes supported at 100\%.

\subsection*{Mammal Phylogeny}

We further tested the utility of SISRS using NGS data from placental mammals, for which the phylogeny is controversial.
This analysis was conducted using 40 cores; the total time to produce the composite genome was less than an hour;
the remaining alignment, base calling, and site identification steps required an additional 
87 hours.
10Gb of memory was required during the composite genome assembly.
The maximum amount of memory required to process the data mapped to the composite genome was 45 Gb; however, because data processing was conducted across multiple cores to increase the speed of the analysis, the total amount of memory used at one time by SISRS was over 300 Gb.
Thus, this analysis could be conducted using fewer resources over more time.

We produced 15 alignments, each allowing a set number of species to be missing data at each site (i.e.\ alignment 1 has no more than one species missing per site). 
The number of sites in the alignments ranged from 21 to 1572322. 
Analyzed in a ML framework, the first four alignments produced phylogenies with multiple polytomies due to limited data; they are not described further.
The remaining alignments produced phylogenies that reflect previous conflicting estimates of the relationships among mammals.
For example, in regards to the basal relationships
alignments 5, 10, 11, and 12 supported Xenarthra+Afrotheria (Atlantogenata) as a clade sister to all other mammals \citep{Meredith2011,Song2012,Morgan2013}, 
alignment 6 supported Xenarthra as a separate clade \citep{OLeary2013}, 
alignments 13, 14, and 15 supported Afrotheria as a separate clade \citep{McCormack2012,Romiguier2013}, 
and the remaining alignments did not resolve these relationships.
Similarly, the relationship of the treeshrew to other mammals was difficult to resolve:
for some alignments this species formed a clade with rodents, while for others it formed a clade with primates.
The majority rule consensus phylogeny generated from alignment 10 (i.e.\ all sites missing data from no more than 10 of the 30 mammal taxa) is shown in Figure \ref{fig:mammal_phyl}; additional phylogenies are shown in Supplementary Figure \ref{fig:more_mammal_phyl}.

\section*{Discussion\label{sec:Discussion}}

The approach introduced here has the potential to transform phylogenetic research.
SISRS eliminates the need for expensive marker development in many studies by using whole genome shotgun sequence data directly.
As technology improves, whole-genome sequencing will soon be affordable even for large-scale projects. 
By using shotgun sequence data, error in next-generation sequence data and co-alignment of paralogous genes does not affect subsequent analyses.

SISRS also promotes the reuse of data.
Shotgun genomic sequences available in public databases can be used directly for phylogenetic analyses, as we have done in this study.
Sequencing performed with the goal of identifying phylogenetic data using SISRS can be made available for subsequent use in other studies, including phylogenetics at any taxonomic level, or any other study utilizing genomic data.
Reusing available next-generation sequencing reads can substantially reduce costs.

\subsection*{Data produced by SISRS}

For simulations, SISRS was able to recover large numbers of variable sites, unless branch lengths were unreasonably long or coverage was low.
Based on these results, coverage should average 5--10x for optimal marker identification; however, low coverage sequencing will also identify useful data.
Most genomes are much larger than the one million bases in our simulations; far more sites will be identified for larger genomes, making the use of low coverage data feasible.
Although site identification was challenging for trees with long branches (i.e.\ large evolutionary distances per number of substitutions per site between taxa), this result is less problematic for empirical data.
In real genomes, unlike our simulations, loci evolve at different rates; thus, there will always be some loci for which the branch length (in substitutions per site) between taxa is very small.
These sites will be identified by SISRS.

As expected, deeper nodes were more difficult to recover, likely because the synapomorphies between the two clades may be overwritten by new substitutions.
However, the sites that were identified are informative about these relationships.
Overall, the concordance of the data with the simulation trees demonstrates that the SISRS approach produces extensive phylogenetically informative data for deep and shallow evolutionary time scales. 

\subsection*{Empirical results}

Using available NGS data we were able to recover the phylogeny of apes quickly and accurately. 
Similarly, using available NGS data we recovered a mammal phylogeny reflecting previous conflicting estimates of the relationships among mammals, particularly in regards to the basal relationships and the position of the treeshrew \citep{Teeling2013,Romiguier2013,Morgan2013,Fan2013}.
We are currently developing metrics to understand where greater phylogenetic signal and noise are found in these datasets.

\subsection*{Advantages of a composite genome}

Generating a composite genome has multiple advantages over aligning data to a reference genome to identify potential phylogenetically informative sites. 
First, an assembled reference genome similar to the taxa of interest is not always available. 
Second, assembling a reference genome requires high coverage data from at least one species and is time consuming; assembling the whole genome is necessary because it is impossible to determine \textit{a priori} regions of the genome that may be phylogenetically informative.
In contrast, SISRS does not require high levels of coverage or a time consuming assembly.
A composite reference genome containing phylogenetically-informative homologous regions can be assembled in a few hours.
Furthermore, when taxa are highly diverged, data may align poorly to a single reference genome. 
In contrast, a composite genome contains data from all taxa, allowing better alignment of all data across the phylogeny. 
Because each species is subsampled for the assembly, unique regions will be limited in the final assembly, while maintaining an optimal assembly for conserved regions. 
Within each conserved region, the composite genome contains sites with the most common base, making it more likely that data from all taxa will align to this region.

\subsection*{Time required}

The time to run SISRS is highly variable, depending on the number of processors available, the number of samples sequenced, and the amount of data sequenced per sample.
Given large numbers of processors (e.g.\ a cluster of $>$30 nodes), SISRS makes phylogenetic analysis from shotgun data possible within a few days.
Even given the limitations of a desktop computer, it is possible to produce many phylogenies within a couple of weeks.
Unlike other phylogenetic methods, SISRS entirely avoids the weeks required for marker development or sample processing; preparation for sequencing and the sequencing time itself are required for all phylogenomic approaches.
Alternatively, as with the analyses conducted here, all time required for sequencing and preparation was avoided entirely by using data made available from other research projects. 

\subsection*{Data analysis}

The approaches we have used to analyze our data are not designed for large datasets of variable sites, although our results suggest that with the exception of short deep internodes the recovery of the phylogeny is quite good.
Ideal methods would accommodate differences among gene trees to correctly estimate the species tree and model the substitution process to accurately infer substitutions. 
However, current methods to analyze genome-wide variable sites are limited \citep{Bryant2012}.
It is important to note that these methods are in development; as we begin to use whole-genome data it is obvious that subsets of data must be used and the optimal data are likely not linked regions, but individual sites \citep{Yoder2013}.
Furthermore, the availability of tools designed for limited datasets should not prevent us from developing methods to identify more comprehensive datasets. 

\subsection*{Future Directions}

SISRS is under active development. 
Future versions will accommodate larger genomes and output more variable sites more rapidly as a result of improved assembly of the composite genome and improved genotype calling.
We will also evaluate the application of SISRS output to deep-time phylogenetics and estimation of branch lengths / divergence times among taxa.

\section*{Acknowledgments}

This work was supported by ASU startup funds to R.\ Cartwright, a National Science Foundation Doctoral Dissertation Improvement Grant [grant number BCS-1232582 to K.\ Harkins and A.\ Stone], a National Institutes of Health Grant [grant number R01-GM101352-01A1 to R.\ Zufall, R.\ Azevedo, and R.\ Cartwright], and a National Science Foundation Advances in Bioinformatics Grant [grant number DBI-1356548 to R.\ Cartwright]. 
Some data were provided by the Wellcome Trust Sanger Institute prior to publication; they can be obtained from the European Nucleotide Archive.
Members of the Cartwright lab provided feedback on this project and manuscript.  

\bibliography{SISRS_ms}
\bibliographystyle{natbib}

\newpage
\newpage
\onecolumn
\section*{Figures}

\begin{figure}[!ht]
\includegraphics[width=1\textwidth]{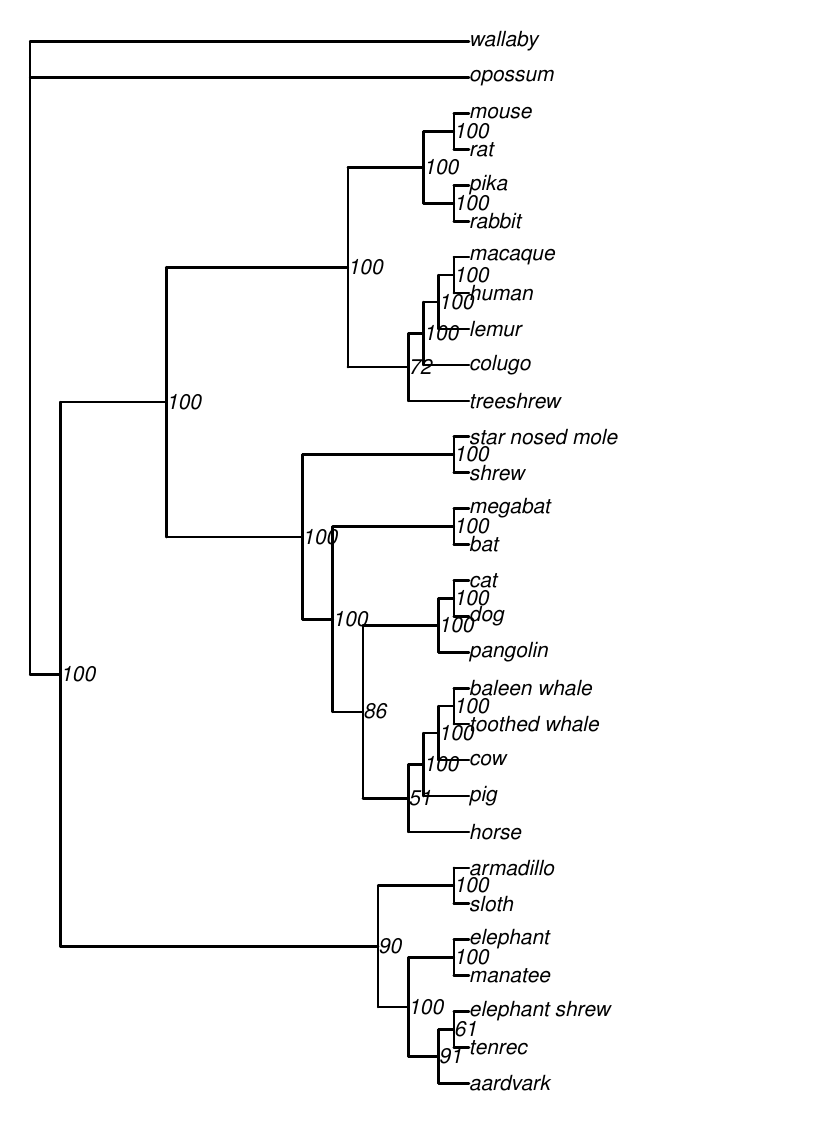}
\caption{The majority rule consensus phylogeny generated with RAxML 8.0 from an alignment of variable sites produced by SISRS with all sites missing data from no more than 10 of the 30 mammal taxa.}
\label{fig:mammal_phyl}
\end{figure}


%

\clearpage
\section*{Supplementary Figures}
\setcounter{figure}{0} 

\begin{figure}[!ht]
\includegraphics[trim=0cm 10cm 0cm 0cm, clip=true, width=\textwidth]{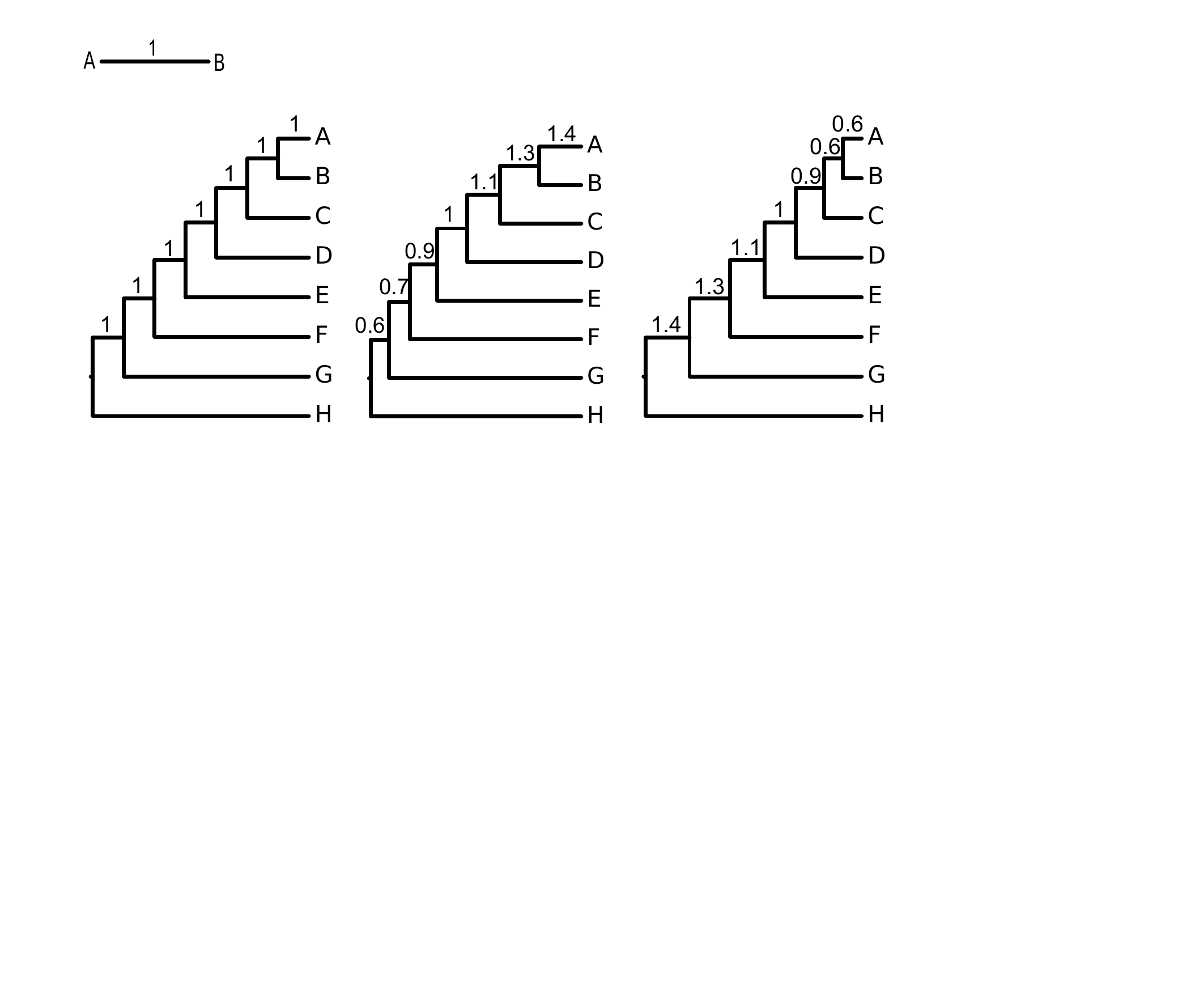}
\caption{Simulation trees used to test the SISRS pipeline. Genomes were simulated on each tree using DAWG 2.0. Branch lengths were scaled by multiplying by 0.01-0.06.}
\label{fig:trees}
\end{figure}

\begin{figure}[!ht]
\includegraphics[width=1\textwidth]{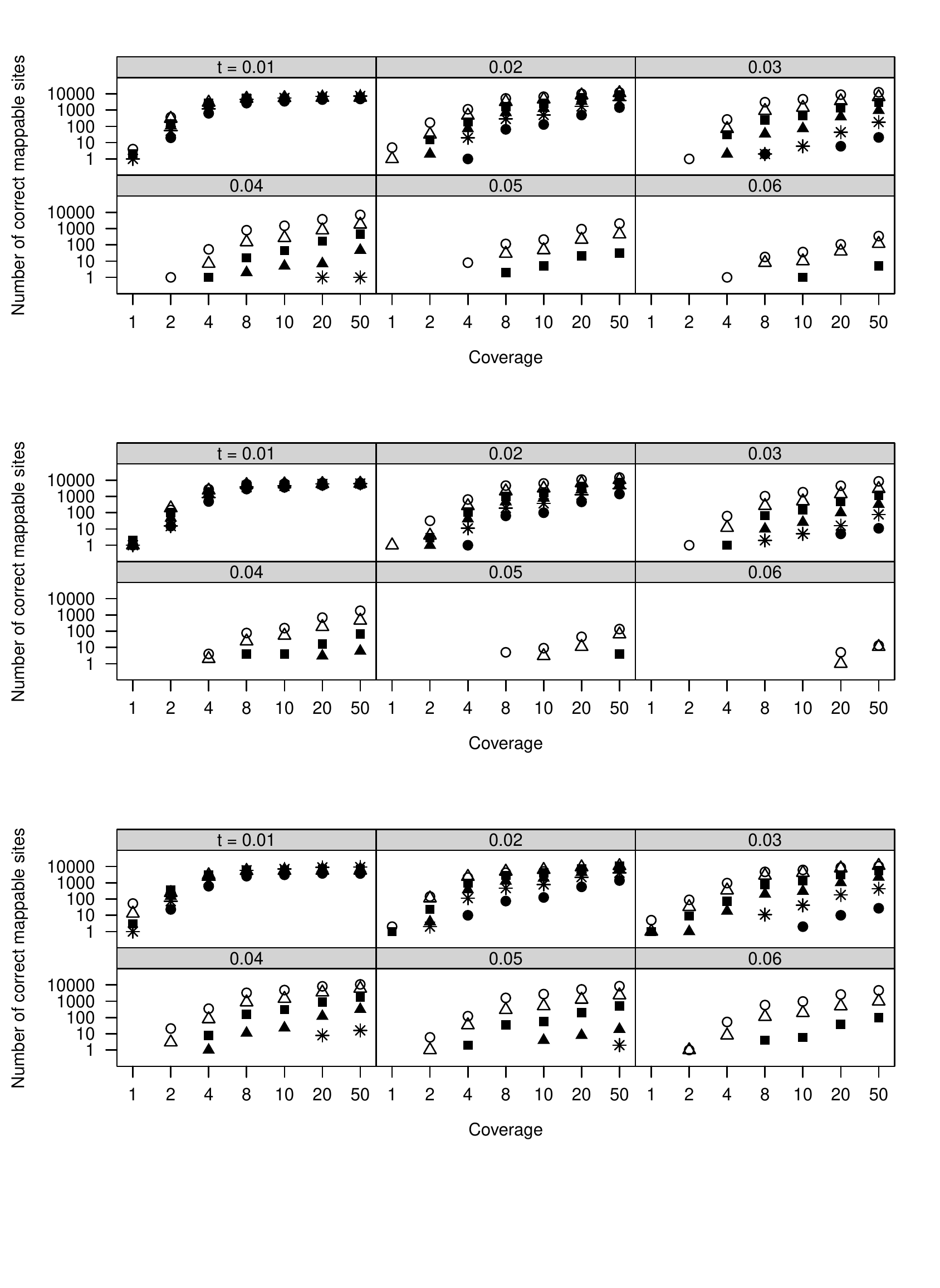}
\caption{SISRS identified fewer informative sites for deeper nodes in the tree; however, in most cases the number of sites was sufficient to resolve the tree. 
Results are separated by branch length (panels), coverage (x axis), and tree depth (symbols). 
The number of sites supporting the node A+B are denoted as $\circ$; $\triangle$ denotes sites supporting A+B+C;$\blacksquare$ denotes A+B+C+D; $\blacktriangle$ denotes A+B+C+D+E; $\ast$ denotes A+B+C+D+E+F; $\bullet$ denotes A+B+C+D+E+F+G.
Top: results for the equal-branch-length tree. Middle: results for the tree with short deep branches. Bottom: results for the tree with long deep branches.}
\label{fig:sim_results_all_depths}
\end{figure}

\begin{figure}[!ht]
\includegraphics[width=1\textwidth]{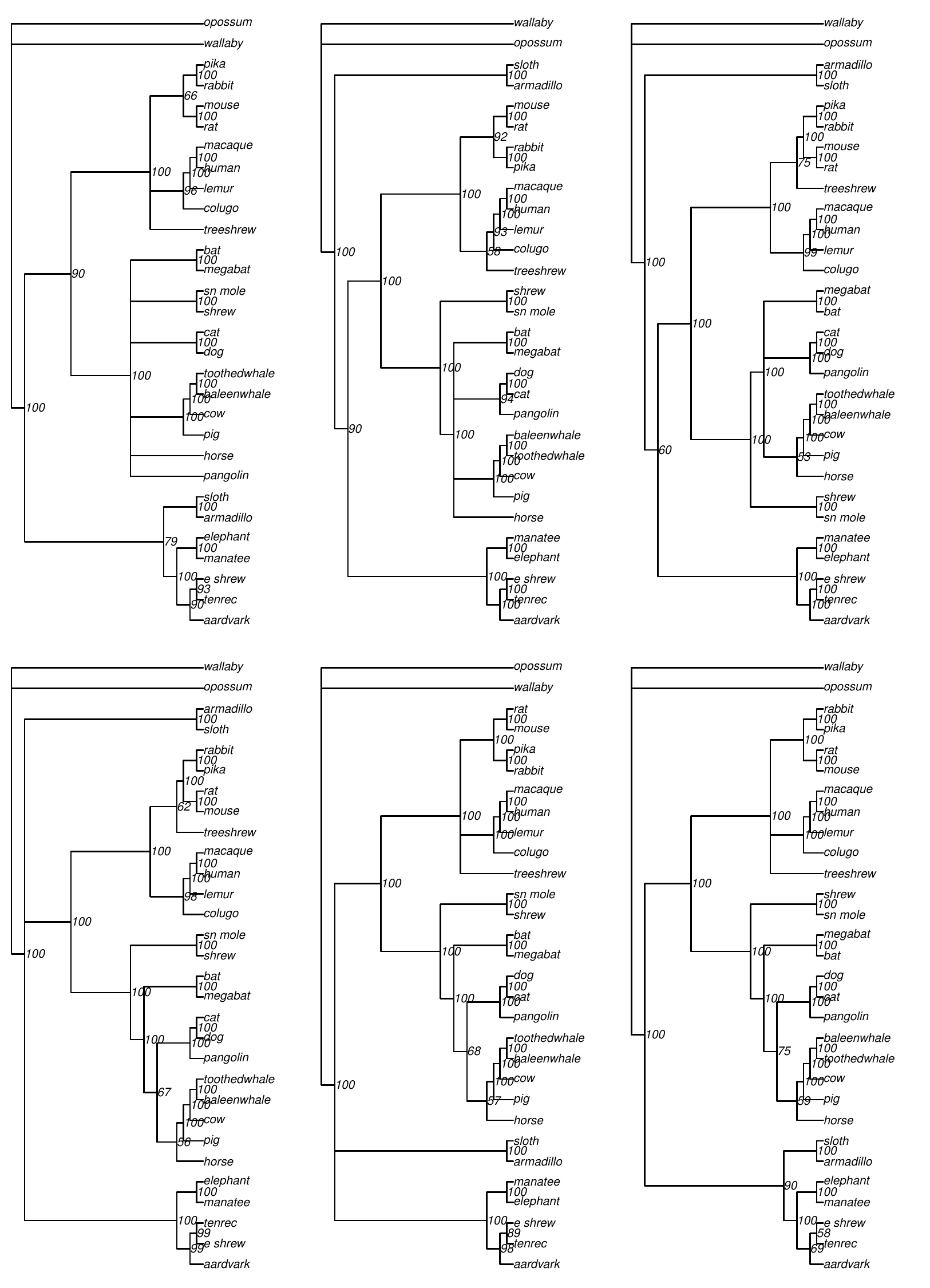}
\caption{Phylogenies estimated for placental mammals using data from SISRS with RAxML 8.0.
The datasets were missing information for up to 5, 6, 7, 8, 9, 11, 12, 13, 14, and 15 species at each site respectively.}
\label{fig:more_mammal_phyl}
\end{figure}


\clearpage
\section*{Supplementary Tables}
\begin{table}[ht]
  \caption{Accession numbers for data downloaded from the European Nucleotide Archive. Human data are from the 1000 genomes project \citep{1000GenomesProjectConsortium2010}. Taxa for which transcriptome data was used are italicized.}
  \centering
  \begin{tabular}{c}
     \includegraphics[width=1\textwidth]{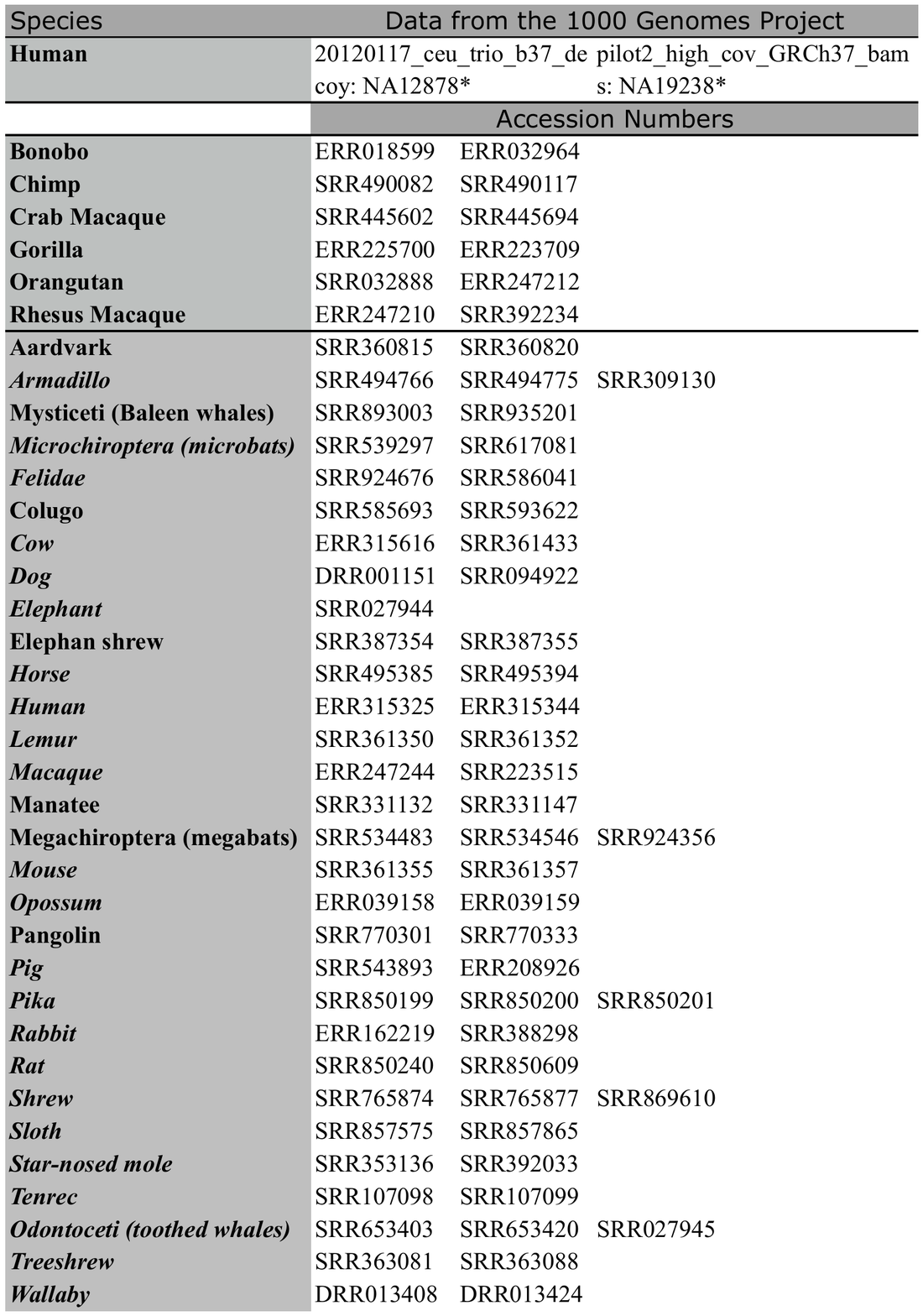}
  \end{tabular}
  \label{table:ENAdata}
\end{table}


\end{document}